\documentstyle[times,pramana,epsf,floats]{ias}

\newcommand{\bm}[1]{\mbox{\boldmath $#1$}}

\def\P{{\bf Proof:} \hspace{3mm}}

\newtheorem{theo}{Theorem}

\newtheorem{prop}{Proposition}
\newtheorem{lem}{Lemma}

\begin{document}
\mark{{Singularity theorem}{J M M Senovilla}}
\title{A singularity theorem  based on spatial averages\footnote{In memoriam Amal Kumar Raychaudhuri (1924-2005)}}

\author{J M M SENOVILLA} \address{F\'{\i}sica Te\'orica, Universidad
del Pa\'{\i}s Vasco, Apartado 644, 48080 Bilbao, Spain}
\keywords{Raychaudhuri equation, singularity-free cosmologies,
singularity theorems} \pacs{04.20.Cv, 04.20.Dw, 98.80.Jk} \abstract{
Inspired by Raychaudhuri's work, and using the equation named after
him as a basic ingredient, a new singularity theorem is proved. Open
non-rotating everywhere expanding universes with non-vanishing {\em
spatial} average of the matter variables are {\em severely}
geodesically incomplete to the past. Another way of stating the result 
is that, under the same conditions, any singularity-free model must
have a vanishing spatial average of the energy density (and other
physical variables). This is very satisfactory and provides a clear
decisive difference between singular and non-singular cosmologies.}

\maketitle

\section{Introduction}
In this paper I would like to present a result which confirms ---at 
least partially--- a 10-year-old conjecture: reasonable non-rotating 
cosmological models can be non-singular only if they are open and 
have vanishing spatial averages of the matter and other physical 
variables. This conjecture arose as a result of the interactions and 
discussions between Professor A K Raychaudhuri (AKR from now on) and 
myself concerning the question of the feasibility of singularity-free 
cosmological solutions, see \cite{R2,S3}. This is a subject to which, as is 
well-known, AKR made fundamental pioneering contributions 
\cite{R,R1}, and I influenced in a much more modest and lateral 
manner \cite{S,S1,S2}. 

The next section contains a more or less detailed historical review of
the antecedents, birth, and   
hazardous life of the conjecture. However, any reader, {\em not} 
interested in this historical appraisal, who only wishes to see and 
learn about the results and the new theorem ---which are interesting 
on their own--- may go directly 
to Sects.\ \ref{Req} and \ref{Th}. The former contains a brief 
derivation of the celebrated Raychaudhuri equation and some 
discussion on the definition of spatial averages; the latter 
presents the new theorem and its implications. A brief discussion 
and an Appendix have been placed at the end of the paper.

\section{History of a conjecture}
This is divided into three parts: the appearance of the first 
non-singular solution and subsequent developments are summarized 
in subsection \ref{AV}; the importance of the fresh new ideas brought 
in by AKR, and the formulation of the conjecture, are analyzed in 
subsection \ref{conj}; finally, the reaction by AKR about the 
conjecture, the sometimes chaotic discussions in 
publications-comments-replies, and some recent developments on the 
subject are reviewed in subsection \ref{mess}.

\subsection{Anadi Vishva}
\label{AV}
In 1990 I published \cite{S} what is considered to be the first 
``interesting'' singularity-free cosmological model. It is a 
spatially inhomogeneous solution of Einstein's field equations for a 
perfect fluid with a realistic equation of state $p=\varrho /3$. This 
solution came as a big surprise: it was widely believed that a model 
such as the one in \cite{S} should be singular due to the powerful 
singularity theorems developed by Penrose, Hawking and others 
\cite{P,HP,HE}. Therefore, the solution caused some impact, see e.g. 
\cite{M}, and some discussions within the 
relativity community \cite{J,S0}.

Shortly afterwards, the geodesic completeness of this solution
was explicitly proven in \cite{CFS}, along with a list of its main
properties. The solution turns out to be {\em 
cylindrically symmetric}, globally hyperbolic and everywhere expanding 
for half the history of the model. It satisfies the stronger energy 
requirements (the dominant and strict strong energy conditions) and 
is singularity-free (see the Appendix for a short summary). A 
detailed analysis of how the model fits in with the general 
conclusions of the singularity theorems was also performed in 
\cite{CFS}, and the solution was proven to be in full accordance 
with the theorems: in all versions of the 
theorems at least one of their hypotheses was not satisfied, usually 
the so-called boundary or initial condition, see \cite{S2}.

Almost simultaneously, another paper \cite{RS} extended the 
particular model in \cite{S} to a general family of 
perfect-fluid solutions with an Abelian $G_{2}$-symmetry acting on 
spatial surfaces. This family was obtained under the assumption 
of separation of variables.  It contains a diversity of models, 
with different properties, in particular a 2-parameter subfamily of 
geodesically complete, singularity-free, solutions ---which include 
the original solution in \cite{S}--- as well as other relevant 
solutions such as the one in \cite{FS}, see also \cite{vBS}. There 
were also solutions with timelike singularities, and solutions with 
no singularities in the matter variables but with timelike 
singularities in the Weyl conformal part of the curvature \cite{RS}. 
All such behaviour was somehow unexpected. As a matter of 
fact, one could prove that, within the subfamily of solutions in 
\cite{RS} with a barotropic equation of state and no initial 
singularity, the singularity-free solutions lie precisely at the 
boundary separating the solutions with only timelike Weyl 
singularities from the solutions with timelike singularities in both 
Weyl and matter variables. This was not very encouraging, since one 
could suspect the {\em instability and the zero measure} of the 
non-singular models.

Several generalizations of these solutions were later found by 
allowing the fluid to have heat flow \cite{PD} and scalar fields 
\cite{DPT}. It was claimed in \cite{DTP,DPT} that a particular 
``inhomogeneisation'' procedure (by requiring the separation of variables) of 
the Robertson-Walker open cosmological models would lead to the 
non-singular models of \cite{S,RS}. This was later pursued in a 
review published by Dadhich in \cite{D}, where he intended to prove 
the uniqueness of those non-singular models.

Up to this point, all known non-singular solutions were (i) {\em 
diagonal}, that is to say, with the existence of a global coordinate chart 
adapted to the perfect fluid (so-called ``co-moving'') such that the 
metric takes a diagonal form, (ii) {\em cylindrically symmetric}, and 
(iii) {\em separable} in comoving coordinates.  The last feature meant
that the metric components 
could be written as the product of a function of the separable time coordinate 
times a function of the separable radial coordinate.
The first example of a 
non-diagonal non-singular perfect fluid model was presented in 
\cite{Mars}, though the solution had been previously published in 
quite a different context in \cite{L}. This was a solution for a 
cylindrically symmetric stiff fluid (the equation of state is 
$p=\varrho$) or equivalently, for a massless scalar field. It was 
also separable but could be almost immediately generalized to a  
family of {\em non-separable}, non-diagonal, non-singular stiff 
fluid solutions in \cite{GB}. 
Other non-singular solutions followed, see e.g. \cite{Mars2,MS,S2}.

It was then shown in \cite{S1} that the general family in \cite{RS} 
as well as the {\em whole} class of Robertson-Walker cosmologies belong to 
a single unified wider class of cylindrically 
symmetric\footnote{Recall that all Robertson-Walker geometries are 
cylindrically symmetric.}, separable, diagonal 
(non-necessarily perfect) fluid solutions. Those depended on one arbitrary 
function of time ---essentially the scale factor--- and four free 
parameters selecting the openness or closeness of the models, the 
anisotropy of the fluid pressures, or the anisotropy and spatial 
inhomogeneity of the models. The physical properties of this 
general class were analyzed in detail \cite{S1,S2}, in particular the 
deceleration parameter, leading to natural inflationary models 
(without violating the strong energy condition), and the generalized 
Hubble law.
The possibility of constructing realistic cosmological models by 
``adiabatically'' changing the parameters in order to start with a 
singularity-free model which at later times becomes a 
Robertson-Walker model was also considered in \cite{S1} and in section 
7.7 in \cite{S2}.

Some interesting lines of research appeared in print in 1997-8. 
First, by keeping the cylindrical symmetry, a new diagonal but 
non-separable family of stiff fluid singularity-free solutions was 
presented in \cite{Leo}. The family contained the same static limit 
as the solution in \cite{Mars}, thereby suggesting  
that they both form part of a more general class, perhaps of non-zero 
measure, of non-singular cylindrically symmetric stiff fluid models. 
Second, the role of shear in expanding perfect fluid models was 
analyzed in \cite{DP} proving that non-singular models with an 
Abelian spatial $G_2$-symmetry should be spatially inhomogeneous. 
This result was much improved and proven in a more general context in 
\cite{B}, showing in particular that the symmetry assumption was 
superfluous. And third, by giving up cylindrical symmetry, a family 
of non-singular (non-perfect) fluid solutions with spherical symmetry was 
presented in \cite{D1}. These models depend on one arbitrary function of 
time. Once again, they can avoid the singularity theorems due to the 
failure of the boundary/initial condition: there are no closed 
trapped surfaces. It was also shown in section 7.8 of \cite{S2} that 
these models cannot represent a finite star, since this would require a 
place where the radial pressure vanished, which is impossible for 
appropriate selections of the arbitrary function of time. This is a 
property shared by all models mentioned so far in this subsection.

\subsection{Raychaudhuri comes into play; the conjecture}
\label{conj}
In December 1995 I attended the International Conference on 
Gravitation and Cosmology (ICGC-95), held in Pune (India), where I 
had the chance to meet Prof. Raychaudhuri for the first time.  
I was impressed by his personality and accessibility, specially for 
a man of his age and reputation. But more importantly, I was deeply 
influenced by his remarks in brief conversations that we ---AKR,
Naresh Dadhich and myself---  
had at that time. If I remember well, AKR mentioned {\em averages} in
these informal conversations, but just by the way.  This came at a
critical time:  in a short talk at the workshop on ``Classical General
Relativity'', see \cite{DM}. I presented the above mentioned combined  
Robertson-Walker plus non-singular general family \cite{S1} and its 
properties. I thought that the paper, which had already been accepted,  
would open the door for realistic models. 

Even though AKR meant {\em spacetime} 
averages, I immediately realized the relevance of his idea, 
especially to discriminate between singular and non-singular 
cosmological models, but using purely {\em spatial} averages.
As remarked at the end of the previous subsection, all known 
non-singular models were ``cosmological'' in the sense that they could 
not describe a finite star surrounded by a surface of vanishing 
pressure. {\em However}, it can certainly happen that (say) the 
energy density falls off too quickly at large distances (this 
certainly occurred in all known singularity-free solutions).   
Thus one may raise the issue whether or not this will better 
describe the actual Universe or rather a weakly-localized object such 
as a very large galaxy. Of course, a good way to distinguish 
between these two possibilities is to use the {\em spatial} average 
of the energy density.
Thus, I was inspired by AKR's remarks and believed that this was the 
right answer to the existence of non-singular models such as the one 
in \cite{S}. I incorporated this view to the review \cite{S2}, see 
p.821. 

I met AKR for the second time in Pune again, on the occasion of the
15th International Conference on General Relativity and Gravitation
(GR15), held in December 1997. Either at GR15 or in an informal
seminar (I cannot exactly recall), I attended a talk where he
discussed some of the  
non-singular models and made some 
comments about the importance of the {\em averages} of the physical 
quantities such as the energy density, the pressure, or the expansion 
of the fluid. 
In 1998 AKR proved \cite{R2} that, under some 
reasonable assumptions, open non-rotating non-singular models must 
have vanishing {\em spacetime} averages of the matter and kinematical 
variables. Later, it was shown   
\cite{Saa,S3} explicitly that, in the open 
Robertson-Walker models (with an initial singularity), the same 
spacetime averages vanish too. Actually, this holds true for most 
open spatially homogeneous models as well. Since this property is 
shared by all models, it cannot be used to decide between 
singularity-free and singular spacetimes.

In my comment, I stressed the following fact. I came to understand,
after listening to  
AKR, that pure spatial averages (at a given instant of time)  
vanish in the known non-singular solutions, while they are 
non-vanishing in open Robertson-Walker models. This, together with a 
well-known singularity theorem for expanding globally hyperbolic 
models (Theorem \ref{th1} below), enabled the formulation of the
following conjecture \cite{S3}: 
\begin{quote}
In every singularity-free, non-rotating, expanding, globally 
hyperbolic model satisfying the strong energy condition, spatial 
averages of the matter variables vanish.
\end{quote}
This will be made precise and proven
in section \ref{Th} below.

\subsection{History of the conjecture}
\label{mess}
My comment \cite{S3} was only submitted after electronic
correspondence with AKR.\footnote{I thank Dr. Narayan Banerjee for
helping with AKR's emails.}  It seems that he was initially skeptical
about the use 
of {\em purely spatial} ---and not spacetime--- averages, for he also
replied in \cite{R3} to our comments.  Nevertheless, in an e-mail
dated on September the 9th, 1998, he mentioned that a letter proving
the vanishing of {\em spatial} averages 
``following his earlier method'' had already been submitted for
publication.  This private announcement was followed, shortly
thereafter, by (i) 
another paper by Dadhich and AKR \cite{DR} where they proved the
existence of oscillatory non-singular models within the non-perfect
fluid spherically symmetric family of \cite{D1} mentioned above, (ii)
general theorems providing sufficient conditions for the geodesic
completeness of general cylindrically symmetric spacetimes
\cite{LeoMan,Leo1} and (iii) some work \cite{G} showing the relevance
that the singularity-free solutions might have in the fashionable
String Cosmology, see also \cite{GV} and references therein.

The letter that AKR mentioned was published in \cite{R4}. However, in
it global hyperbolicity was assumed without being mentioned (due to
the assumption of the existence of global coordinates associated to a 
hypersurface-orthogonal timelike eigenvector field of the Ricci
tensor).  Moreover, the openness of some local  
coordinates was taken for granted. Further, the statement that the
spatial average 
of the divergence of the acceleration associated to the timelike 
eigenvector field vanishes was not clearly proved. More importantly, the 
blow-up of the kinematical quantities of this eigenvector field was 
incorrectly related to the blow up of some Ricci 
scalar invariants.\footnote{Blowing-up of the expansion of the Ricci timelike eigendirection can take place due to many different reasons such as a caustic, 
the failure of hypersurface orthogonality or of local coordinates, 
the failure of the vector field to 
be a Ricci eigendirection, or a change in the algebraic type of
the Ricci tensor, among others. There are well-known examples of
hypersurface-orthogonal  timelike congruences showing this behaviour even in flat spacetime.} All in all, the result in 
\cite{R4}, involving spatial averages, was not completely proved.

This led to some interesting works by others   
\cite{LeoMan1,LeoMan2}, where the existence of a wide class of
singularity-free  
(geodesically complete) cylindrically symmetric stiff fluid cosmologies 
was {\em explicitly} demonstrated, and many solutions were actually 
exhibited. In these papers, the family 
of regular cylindrically symmetric stiff fluids was proven to be very
abundant, allowing for arbitrary functions. Furthermore,
strong support for the conjecture was also provided in the second of these papers \cite{LeoMan2}: the vanishing 
of the energy density (and pressure) of the fluid at spatial infinity 
on every Cauchy hypersurface was demonstrated to be a necessary 
requirement if the spacetime was to be geodesically complete. This was 
quite encouraging, and constituted the first serious advance towards 
the proof of the conjecture.

AKR may not have been aware of these important developments and 
results. He put out a preprint \cite{Rgrqc} to which L. 
Fern\'andez-Jambrina \cite{Leogrqc} found a counter-example. 
A revised version was published in \cite{R5} which was also not fully
free of the shortcomings, as neatly pointed out in \cite{Leo2}. AKR
acknowledged these deficiencies in \cite{R6}, yet however he believed
that some of his results were still valid. 

Ref. \cite{R6} was AKR's last published paper. In my opinion, after
having identified the clue to non-singular models (i.e. averages),
which led us all to the right track, he tried to prove more ambitious and
challenging results which were perhaps beyond the techniques he was
using. This, of course, does not in any way diminishes his fundamental
contribution to the field of singularities in Cosmology, a subject in
which, probably, the most important ideas came from his insight and
deep intuition. It is in this sense that the spirit of the theorem to
be proven in section \ref{Th} should be credited to him. I can only
hope that he would have welcomed the new results.

\section{Spatial averages and the Raychaudhuri equation}
\label{Req}
\subsection{The Raychaudhuri equation}
As is known, the first result predicting singularities under reasonable physical 
conditions was published in 1955 ---exactly the year of 
Einstein's demise--- by AKR \cite{R}. In this remarkable paper, he 
presented what is considered to be the first singularity theorem, and 
included a version (the full equation appeared soon after in 
\cite{K}, see also \cite{R1}) of the equation named after him which 
is the basis of later developments and of
{\em all} the singularity theorems \cite{HE,HP,P,S2}. The 
Raychaudhuri 
equation can be easily derived from the general 
Ricci identity:
$$
(\nabla_{\mu}\nabla_{\nu}-\nabla_{\nu}\nabla_{\mu})u^{\alpha}=
R^{\alpha}{}_{\rho\mu\nu}u^{\rho}\, .
$$
Contracting $\alpha$ with $\mu$ here, then with $u^{\nu}$, one gets
$$
u^{\nu}\nabla_{\mu}\nabla_{\nu}u^{\mu}-u^{\nu}\nabla_{\nu}\nabla_{\mu}u^{\mu}=
R_{\rho\nu}u^{\rho}u^{\nu}
$$
where $R_{\mu\nu}$ is the Ricci tensor.
Reorganizing by parts the 
first summand on the left-hand side, one derives
\begin{equation}
\fbox{$\displaystyle{
u^{\nu}\nabla_{\nu}\nabla_{\mu}u^{\mu}+\nabla_{\mu}u_{\nu}\nabla^{\nu}u^{\mu}-
\nabla_{\mu}(u^{\nu}\nabla_{\nu}u^{\mu})+R_{\rho\nu}u^{\rho}u^{\nu}=0
}$}
\label{rayeq}
\end{equation}
which is the Raychaudhuri equation. AKR's important 
contribution was to understand and explicitly show the fundamental
physical implications of this simple geometrical relation. 

Let us analyze some of these implications. Observe that in the case
that $u^\mu$ defines a 
(affinely parametrized) {\em geodesic} vector field, then 
$u^{\nu}\nabla_{\nu}u^{\mu}=0$ and the third term vanishes. The 
second term can then be rewritten by splitting 
$$
\nabla_{\mu}u_{\nu}=S_{\mu\nu}+A_{\mu\nu}
$$
into its symmetric $S_{\mu\nu}$ and antisymmetric $A_{\mu\nu}$ parts, 
so that
$$
\nabla_{\mu}u_{\nu}\nabla^{\nu}u^{\mu}=S_{\mu\nu}S^{\mu\nu}-A_{\mu\nu}A^{\mu\nu}\, .
$$
Now the point is to realize two things.  (i) If $u^{\mu}$ is timelike (and 
normalized) or null, then both $S_{\mu\nu}S^{\mu\nu}$ and 
$A_{\mu\nu}A^{\mu\nu}$ are non-negative. (ii) $u_{\mu}$ is also
proportional to a gradient (therefore defining orthogonal 
hypersurfaces) if and only if $A_{\mu\nu}=0$. In summary, for 
hypersurface-orthogonal geodesic time-like or null vector fields 
$u^{\mu}$, one has
$$
u^{\nu}\nabla_{\nu}\nabla_{\mu}u^{\mu}=
-S_{\mu\nu}S^{\mu\nu}-R_{\rho\nu}u^{\rho}u^{\nu},
$$
so that the sign of the derivative of the divergence or {\em expansion}
$\theta\equiv \nabla_{\mu}u^{\mu}$ along the geodesic congruence
is governed by the sign 
of $R_{\rho\nu}u^{\rho}u^{\nu}$. If the latter is non-negative, then 
the former is non-positive. In particular, if the expansion is 
negative at some point and $R_{\rho\nu}u^{\rho}u^{\nu}\geq 0$ then one 
can prove, by introducing a scale factor $L$ such that 
$u^{\mu}\nabla_{\mu}(\log L) \propto \theta$ and noting 
that $S^{\mu}{}_{\mu}=\theta$, that
necessarily the divergence will reach an infinite negative value in 
finite affine parameter (unless all the quantities are zero 
everywhere).

If there are physical particles moving along these geodesics, then 
clearly a physical singularity is obtained, since the mean volume 
decreases and the density of particles will be unbounded, see Theorem
5.1 in \cite{S2}, p.787. This was  
the situation treated by AKR for the case of irrotational 
dust. In general, no singularity is predicted, though, and one only 
gets a typical {\em caustic} along the flow lines of the congruence 
defined by $u^{\mu}$. This generic property is usually called the 
{\em focusing effect} on causal geodesics. For this to take place, of 
course, one needs the condition 
\begin{equation}
R_{\rho\nu}u^{\rho}u^{\nu}\geq 0 \label{sec}
\end{equation}
which is a {\em geometric} condition and independent of the 
particular theory. However, in General Relativity, one can relate the 
Ricci tensor to the energy-momentum tensor $T_{\mu\nu}$ via 
Einstein's field equations 
($8\pi G=c=1$)
\begin{equation}
R_{\mu\nu}-\frac{1}{2}g_{\mu\nu}R+\Lambda g_{\mu\nu}=T_{\mu\nu}\, .
\label{efe}
\end{equation}
Here $R$ is the scalar curvature and $\Lambda$ the 
cosmological constant. Thereby, the condition (\ref{sec}) can be 
rewritten in terms of physical quantities. This is why sometimes 
(\ref{sec}), when valid for all time-like $u^{\mu}$, is called the {\em timelike convergence condition}, also the {\em strong energy condition} in the case with $\Lambda =0$ \cite{HE}. One should bear in mind, 
however, that this is a condition on the Ricci tensor (a geometrical 
object) and therefore will not always hold: see the discussion in 
sect.~6.2 in \cite{S2}.

The focusing effect on causal geodesics predicted by the Raychaudhuri 
equation was the fundamental ingredient needed to derive the 
powerful singularity theorems. Nevertheless, as remarked above, this 
focusing does not lead to singularities on its own in general. As a 
trivial example, observe that flat spacetime satisfies the condition 
(\ref{sec}) trivially, but there are no singularities: the focusing 
effect simply leads to focal points or caustics of the geodesic 
congruences. This is why it took some time to understand the necessity 
of combining the focusing effect with the theory of existence of 
geodesics maximizing the interval (which necessarily cannot have 
focal points or caustics) in order to prove results on {\em geodesic 
incompleteness}, which is a {\em sufficient} condition in the 
accepted definition of a singularity. With the imaginative and 
fruitful ideas put forward by Penrose in the 1960s, later 
followed by Hawking, this led to the celebrated singularity theorems, 
see \cite{P,HP,HE,S2}.

As a simple but powerful example, which we shall later need to prove 
the new theorem, let us present the following standard singularity 
theorem (Theorem 9.5.1 in \cite{Wald}, Theorem 5.2 in \cite{S2}.)
\begin{theo}
If there is a Cauchy hypersurface $\Sigma$ such that the timelike
geodesic congruence emanating orthogonal to $\Sigma$ has an initial expansion
$\theta |_{\Sigma}\geq b>0$ and (\ref{sec}) holds along the congruence, 
then all timelike geodesics are past incomplete.\label{th1}
\end{theo}
The idea of the proof is simple \cite{HE,S2,Wald}: since $\Sigma$ is a Cauchy 
hypersurface, the spacetime is globally hyperbolic so that
one knows that there is a maximal timelike curve from $\Sigma$ to any point. 
From standard results any such maximal curve must be a timelike geodesic 
orthogonal to $\Sigma$ {\em without} any point focal to $\Sigma$
between $\Sigma$ and the point. But the Raychaudhuri equation 
(\ref{rayeq}) implies 
that these focal points should exist to the past at a proper time less 
than or equal to a fixed value (given by $3/\theta |_{\Sigma} \leq 3/b$.) As every causal curve crosses the Cauchy hypersurface $\Sigma$, no timelike geodesic can have length greater than $3/b$ to the past. 

\subsection{Spatial averages}
Let $\Sigma$ be any spacelike hypersurface in the spacetime and let 
$\bm{\eta}_{\Sigma}$ be the canonical volume element 3-form on 
$\Sigma$. The average $\left<f\right>_S$ of any scalar $f$ on a 
finite portion $S$ of $\Sigma$ is defined by
$$
 \left<f\right>_{S}\equiv 
 \frac{\displaystyle{\int_{S}f\bm{\eta}_{\Sigma}}}
{\displaystyle{\int_{S}\bm{\eta}_{\Sigma}}}=
[{\rm Vol}(S)]^{-1}\int_{S}f\bm{\eta}_{\Sigma}
$$
where ${\rm Vol}(S)$ is the volume of $S\subseteq \Sigma$. 
The spatial average on the whole $\Sigma$ is defined as (the limit 
of) the previous expression when $S$ approaches the entire $\Sigma$
\begin{equation}
 \left<f\right>_{\Sigma}\equiv 
 \lim_{S\rightarrow \Sigma}
 \frac{\displaystyle{\int_{S}f\bm{\eta}_{\Sigma}}}
{\displaystyle{\int_{S}\bm{\eta}_{\Sigma}}} \equiv
[{\rm Vol}(\Sigma)]^{-1}\int_{\Sigma}f\bm{\eta}_{\Sigma}
\label{average}
\end{equation}
Obvious properties of these averages are
\begin{enumerate}
\item(linearity) For any $S\subseteq \Sigma$, any functions $f,g$ and 
any constants $a,b$:
$$
\left<af+bg\right>_{S}=a\left<f\right>_{S}+b\left<g\right>_{S}
$$
\label{i}
\item For any $S\subseteq \Sigma$, $\left<f\right>_{S}\leq 
\left<|f|\right>_{S}$.
\label{ii}
\item If $S\subseteq \Sigma$ is such that its closure is compact
(so that its volume is finite Vol$(S)< \infty$), 
then for any $f\geq b\geq 0$ on $S$, $\left<f\right>_{S}\geq b$ 
and the equality holds only if $f$ is constant, $f=b$, almost everywhere 
on $S$. In particular, if $f\geq 0$ on such an $S$, then 
$\left<f\right>_{S}\geq 0$ and the equality holds only if $f$ 
vanishes almost everywhere on $S$.
\label{iii}
\item If $S\subseteq \Sigma$ does not have a finite volume (so that 
it cannot be of compact closure), then for any $f\geq 0$ on $S$, 
$\left<f\right>_{S}\geq 0$ and the equality 
requires necessarily that $f\rightarrow 0$ when ``approaching the boundary'' (i.e., 
when going to infinity). Conversely, if $f> 0$, $f$ is bounded on $S$ 
and $f$ is bounded from below by a positive constant at most along a set of directions of zero measure, 
then $\left<f\right>_{S}= 0$.
\label{iv}
\item If $|f|$ is bounded on $S$ ($|f|\leq M$), then 
$\left<f^2\right>_{S}\leq M\left<|f|\right>_{S}$. In particular, 
$\left<|f|\right>_{S} =0$ implies that $\left<f^2\right>_{S}=0$.
\label{v}
\item Similar results hold, of course, for negative and non-positive 
functions (just use $-f$.)
\label{vi}
\end{enumerate}

\section{The theorem}
\label{Th}
Let us consider any spacelike hypersurface $\Sigma$ in the spacetime, and let 
$u^{\mu}$ be its unit normal vector field (ergo timelike). The 
projector $h_{\mu\nu}=g_{\mu\nu}+u_{\mu}u_{\nu}$ defines the 
canonical first fundamental form of $\Sigma$. A 
classical result relates the intrinsic Riemannian structure of 
$(\Sigma ,h_{\mu\nu})$ with the Lorentzian one of the spacetime, and in 
particular their respective curvatures. The main result is the Gauss equation,
which reads (e.g.\cite{HE,Wald})
$$
R_{\alpha\beta\gamma\delta}
h^{\alpha}_{\lambda}h^{\beta}_{\mu}h^{\gamma}_{\nu}h^{\delta}_{\tau}=
\overline{R}_{\lambda\mu\nu\tau}+K_{\lambda\nu}K_{\mu\tau}-
K_{\lambda\tau}K_{\mu\nu}
$$
where $K_{\mu\nu}=K_{\nu\mu}\equiv 
h^{\beta}_{\mu}h^{\gamma}_{\nu}\nabla_{\beta}u_{\gamma}$ is the second 
fundamental form of $\Sigma$, and 
$\overline{R}_{\lambda\mu\nu\tau}$ is the intrinsic curvature tensor of 
$(\Sigma,h_{\mu\nu})$. Observe that $u^{\mu}K_{\mu\nu}=0$ and 
$u^{\mu}\overline{R}_{\lambda\mu\nu\tau}=0$. Contracting all indices 
here one derives the standard result (e.g. section 10.2 in \cite{Wald})
$$
    K^2=K_{\mu\nu}K^{\mu\nu}+2R_{\mu\nu}u^{\mu}u^{\nu}+R-\overline{R}
$$
where $\overline{R}$ is the scalar curvature of $\Sigma$ and $K\equiv 
K^{\mu}{}_{\mu}$. Using 
the Einstein field equations (\ref{efe}) this can be rewritten in 
terms of the energy-momentum tensor as
\begin{equation}
    K^2=K_{\mu\nu}K^{\mu\nu}+2T_{\mu\nu}u^{\mu}u^{\nu}+2\Lambda -\overline{R}.
    \label{K2}
\end{equation}
Recall that $T_{\mu\nu}u^{\mu}u^{\nu}$ is the energy density of the 
matter content relative to the observer $u^{\mu}$, and thus it is 
always non-negative. Note, also, that for {\em any} extension of 
$u^{\mu}$ outside $\Sigma$ as a hypersurface-orthogonal unit timelike vector 
field (still called $u^{\mu}$), their previously defined kinematical quantities
$\theta=\nabla_{\mu}u^{\mu}$ and $S_{\mu\nu}=\nabla_{(\mu}u_{\nu)}$ are simply
$$
\theta|_{\Sigma}=K, \,\,\,\, \left.\left(S_{\mu\nu}+a_{(\mu}u_{\nu)}\right)\right|_{\Sigma}=K_{\mu\nu}
$$
where $a^{\mu}=u^{\nu}\nabla_{\nu}u^{\mu}$ is the acceleration vector field of the extended $u^{\mu}$.

Combining this with Theorem \ref{th1} one immediately deduces a very 
strong result concerning the average of the energy density:
\begin{prop}
Assume that
\begin{enumerate}
\item there is a non-compact Cauchy hypersurface $\Sigma$ such that the timelike
geodesic congruence emanating orthogonal to $\Sigma$ is expanding and (\ref{sec}) 
holds along the congruence,
\item the spatial scalar curvature is non-positive on average on 
$\Sigma$: $\left<\overline{R}\right>_{\Sigma}\leq 0$,
\item the cosmological constant is non-negative $\Lambda \geq 0$,
\item the spacetime is past timelike geodesically complete.
\end{enumerate}
Then, 
\begin{equation}
\left<T_{\mu\nu}u^{\mu}u^{\nu}\right>_{\Sigma}=
\left<K_{\mu\nu}K^{\mu\nu}\right>_{\Sigma}=\left<\overline{R}\right>_{\Sigma}=
\Lambda=0 \, .\label{cosa}
\end{equation}
\label{th2}
\end{prop}
{\bf Remarks:} 
\begin{itemize}
\item The first assumption requires  
spacetime to be globally hyperbolic (so that it is causally 
well-behaved). Also, the timelike convergence condition has to hold along the 
geodesic congruence orthogonal to one of the Cauchy hypersurfaces. 
Furthermore, the Universe is assumed to be non-closed 
(non-compactness) and everywhere expanding 
at a given instant of time ---described by the hypersurface $\Sigma$---. 
All this is standard, cf. Theorem \ref{th1}.
\item The second assumption demands that the space of the Universe (at 
the expanding instant) be non-positively curved {\em on average}. 
Observe that this still allows for an everywhere positively curved $\Sigma$ (see the example in the Appendix). This 
is in accordance with our present knowledge of the Universe and with 
all indirect observations and measures, e.g. the recent data 
\cite{wmap} from WMAP.
\item The third assumption is also in accordance with all theoretical and
observational data \cite{wmap}. Observe that the traditional case with 
$\Lambda =0$ is included. Notice, however, that 
the timelike convergence condition (\ref{sec}) is also assumed. This
may impose  
very strict restrictions on the matter variables if $\Lambda >0$. 
\item The second and third assumptions could be replaced by milder ones 
such as $\left<\overline{R}\right>_{\Sigma}\leq 2\Lambda$, allowing 
for all signs in both $\Lambda$ and $\left<\overline{R}\right>_{\Sigma}$. 
The conclusion concerning the vanishing of the averaged energy density 
would be unaltered, as well as the next one in (\ref{cosa}), 
but the last two equalities in (\ref{cosa}) should be replaced by
$\left<\overline{R}\right>_{\Sigma}= 2\Lambda$.
\item Probably, the fourth condition may be relaxed substantially, as one just needs
that the set of past-incomplete geodesics be not too big. There are some technical difficulties, however, to find a precise formulation of the mildest acceptable condition.
\item The conclusion forces the second and third assumptions to hold 
in the extreme cases and, much more importantly, it implies that the 
{\em energy density of the matter on $\Sigma$ has a vanishing spatial 
average}. This conclusion was the main goal in this paper. 
\end{itemize}
\P
From theorem \ref{th1} and the fourth hypothesis it follows that 
$0< \theta |_{\Sigma}=K$ cannot be bounded from below by a positive 
constant. Furthermore, the existence of a {\em complete} maximal timelike curve (which must be a geodesic without focal points) from the Cauchy hypersurface $\Sigma$ to any point to the past implies that, actually, $K$ can be bounded from below away from zero only along a set of directions of zero measure. Point \ref{iv} of the list of properties for averages 
implies that $\left<\theta\right>_{\Sigma}=\left<K\right>_{\Sigma}=0$, 
and point \ref{v} in 
the same list provides then $\left<K^2\right>_{\Sigma}=0$. Taking 
averages on formula (\ref{K2}) and using point \ref{i} in that list one 
arrives at
$$
\left<K_{\mu\nu}K^{\mu\nu}\right>_{\Sigma}+
2\left<T_{\mu\nu}u^{\mu}u^{\nu}\right>_{\Sigma}+2\Lambda 
-\left<\overline{R}\right>_{\Sigma}=0.
$$
Then, given that all the summands here are non-negative, the result 
follows.

This result can be made much stronger by using, once again, the Raychaudhuri equation. To that end, we need a Lemma first.
\begin{lem}
If the energy-momentum tensor satisfies the dominant energy condition and 
$\left<T_{\mu\nu}u^{\mu}u^{\nu}\right>_{\Sigma}=0$ for some unit timelike vector field $u^{\mu}$, then
\underline{all} the components of $T_{\mu\nu}$ in \underline{any} orthonormal basis $\{e^{\mu}_{\alpha}\}$ have vanishing average on $\Sigma$
\begin{equation}
\left<T_{\mu\nu}e^{\mu}_{\alpha}e^{\nu}_{\beta}\right>_{\Sigma}=
0 \hspace{1cm}
\forall \alpha,\beta =0,1,2,3.
\label{cosa2}
\end{equation}
\label{lema}
\end{lem}
\P The dominant energy condition implies that \cite{HE,s-e}, in any orthonormal basis $\{e^{\mu}_{\alpha}\}$ (where $e^{\mu}_0$ is the timelike leg)
$$
T_{\mu\nu}e^{\mu}_{0}e^{\nu}_{0}\geq
\left|T_{\mu\nu}e^{\mu}_{\alpha}e^{\nu}_{\beta}\right|, \hspace{1cm}
\forall \alpha,\beta =0,1,2,3, 
$$
so that, by taking any orthonormal basis with $u^{\mu}=e^{\mu}_0$,
points \ref{ii} and \ref{vi} in the list of properties of spatial
averages lead to (\ref{cosa2}) in those bases. As any other
orthonormal basis is obtained from the selected one by means of a
Lorentz transformation ---so that the components of $T_{\mu\nu}$ in
the new basis are linear combinations, with bounded coefficients, of
the original ones--- the result follows. 

The combination of this lemma with proposition \ref{th2} leads to the following result.
\begin{prop}
Under the same assumptions as in proposition \ref{th2}, if $T_{\mu\nu}$ satisfies the dominant energy condition then not only the averages shown in (\ref{cosa}) and (\ref{cosa2}) vanish, but furthermore
\begin{equation}
\left<u^{\mu}\nabla_{\mu}\theta-\nabla_{\mu}a^{\mu}\right>_{\Sigma}=
\left<v^{\mu}\nabla_{\mu}\vartheta\right>_{\Sigma}=
\left<R_{\mu\nu}e^{\mu}_{\alpha}e^{\nu}_{\beta}\right>_{\Sigma}=0,
\label{cosa3}
\end{equation}
where $v^{\mu}$ is the unit timelike geodesic vector field orthogonal to $\Sigma$, $\vartheta=\nabla_{\mu}v^{\mu}$ its expansion, $u^{\mu}$ is any hypersurface-orthogonal  unit timelike vector field orthogonal to $\Sigma$, and $a^{\mu}=u^{\nu}\nabla_{\nu}u^{\mu}$ its acceleration.
\label{th3}
\end{prop}
{\bf Remarks:} 
\begin{itemize}
\item The hypersurface-orthogonal vector field $u^{\mu}$ may represent
  the mean motion of the matter content of the Universe. Observe that,
  thereby, an acceleration of the cosmological fluid is
  permitted. This is important, since such an acceleration is related to
  the existence of pressure gradients, and these forces oppose
  gravitational attraction. 
\item The expression $u^{\mu}\nabla_{\mu}\theta$ can be seen as a ``time derivative" of the expansion $\theta$: a derivative on the transversal direction to $\Sigma$. In particular, $v^{\mu}\nabla_{\mu}\vartheta$ is the time derivative, with respect to proper time, of the expansion for the {\em geodesic} congruence orthogonal to $\Sigma$. This proper time derivative of $\vartheta$ does have a vanishing average on $\Sigma$. Notice, however, that the generic time derivative of the expansion $\theta$ does not have a vanishing average in general: this is governed by the average of the divergence of the {\em acceleration}.
\item Evidently, $a^{\mu}u_{\mu}=0$ so that $a^{\mu}$ is spacelike and
  tangent to $\Sigma$ on $\Sigma$. Furthermore, one can write  
$$
\nabla_{\mu}a^{\mu}=(h^{\mu\nu}-u^{\mu}u^{\nu})\nabla_{\mu}a_{\nu}=
h^{\mu\nu}\nabla_{\mu}a_{\nu}+a_{\mu}a^{\mu}\, .
$$
Letting $\vec a$ represent the spatial vector field $a^{\mu}|_{\Sigma}$ on $\Sigma$, this implies
$$
\nabla_{\mu}a^{\mu}|_{\Sigma}=div_{\Sigma}\vec a+\vec a \cdot \vec a,
$$
where $div_\Sigma$ stands for the 3-dimensional divergence within $(\Sigma,h_{\mu\nu})$ and $\cdot$ is its internal positive-definite scalar product; hence $\vec a \cdot \vec a\geq 0$ and this vanishes only if $\vec a=\vec 0$. Note that the average $< div_{\Sigma}\vec a >_{\Sigma}$ will vanish for any reasonable behaviour of $a^{\mu}|_{\Sigma}$, because the integral in the numerator leads via Gauss theorem to a boundary (surface) integral ``at infinity", which will always be either finite or with a lower-order divergence than $vol(\Sigma)$. Therefore, by taking averages of the previous expression one deduces
$$
\left<\nabla_{\mu}a^{\mu}\right>_{\Sigma}=\left<\vec a \cdot \vec a\right>_{\Sigma}=
\left<a_{\nu}a^{\nu}\right>_{\Sigma}\geq 0.
$$
In other words, the first conclusion in (\ref{cosa3}) can be rewritten in a more interesting way as $\left<u^{\mu}\nabla_{\mu}\theta-a_{\nu}a^{\nu}\right>_{\Sigma}=0$, or equivalently
$$
\left<u^{\mu}\nabla_{\mu}\theta\right>_{\Sigma}=\left<a_{\nu}a^{\nu}\right>_{\Sigma}\geq 0.
$$
Observe that, if these averages do not vanish, this implies in particular that there must be regions on $\Sigma$ where $u^{\mu}\nabla_{\mu}\theta$ is positive.
\end{itemize}
\P
From proposition \ref{th2} it follows that necessarily $\Lambda =0$,
hence in any orthonormal basis $\{e^{\mu}_{\alpha}\}$ the equations (\ref{efe}) imply
$$
R_{\mu\nu}e^{\mu}_{\alpha}e^{\nu}_{\beta}=T_{\mu\nu}e^{\mu}_{\alpha}e^{\nu}_{\beta}-
\frac{1}{2}T^{\rho}{}_{\rho}\eta_{\alpha\beta},
$$
where $(\eta_{\alpha\beta})=$diag $(-1,1,1,1)$. By taking averages on $\Sigma$ here and using lemma \ref{lema} one deduces
$$
\left<R_{\mu\nu}e^{\mu}_{\alpha}e^{\nu}_{\beta}\right>_{\Sigma}=0 \hspace{1cm}
\forall \alpha,\beta =0,1,2,3
$$
which are the last expressions in (\ref{cosa3}).

The Raychaudhuri equations (\ref{rayeq}) for $u^{\mu}$ and $v^{\mu}$ are respectively
\begin{eqnarray}
u^{\nu}\nabla_{\nu}\theta+\nabla_{\mu}u_{\nu}\nabla^{\nu}u^{\mu}-
\nabla_{\mu}a^{\mu}+R_{\mu\nu}u^{\mu}u^{\nu}=0,\label{rayu}\\
v^{\nu}\nabla_{\nu}\vartheta+\nabla_{\mu}v_{\nu}\nabla^{\nu}v^{\mu}+R_{\mu\nu}v^{\mu}v^{\nu}=0.
\label{rayv}
\end{eqnarray}
Obviously $u^{\mu}|_{\Sigma}=v^{\mu}|_{\Sigma}$ and it is elementary to check that $\theta|_{\Sigma}=\vartheta|_{\Sigma}$ and
$$
\nabla_{\mu}u_{\nu}\nabla^{\nu}u^{\mu}|_{\Sigma}=\nabla_{\mu}v_{\nu}\nabla^{\nu}v^{\mu}|_{\Sigma}=
K_{\mu\nu}K^{\mu\nu}.
$$
From (\ref{rayu}-\ref{rayv}) one thus gets
$$
\left.\left(u^{\nu}\nabla_{\nu}\theta- \nabla_{\mu}a^{\mu}\right)\right|_{\Sigma}=
\left.v^{\nu}\nabla_{\nu}\vartheta\right|_{\Sigma}=-K_{\mu\nu}K^{\mu\nu}-
\left.R_{\mu\nu}u^{\mu}u^{\nu}\right|_{\Sigma}\leq 0
$$
so that $u^{\nu}\nabla_{\nu}\theta- \nabla_{\mu}a^{\mu}$ is everywhere non-positive on $\Sigma$. Taking averages on $\Sigma$, using the second in (\ref{cosa}) and the previous result on the averages of the Ricci tensor components one finally gets
$$
\left.\left<u^{\nu}\nabla_{\nu}\theta- \nabla_{\mu}a^{\mu}\right>\right|_{\Sigma}=
\left.\left<v^{\nu}\nabla_{\nu}\vartheta\right>\right|_{\Sigma}=0.
$$
That ends the proof.

The main theorem in this paper is now an immediate corollary of the previous propositions.
\begin{theo}
Assume that
\begin{enumerate}
\item there is a non-compact Cauchy hypersurface $\Sigma$ such that the timelike
geodesic congruence emanating orthogonal to $\Sigma$ is expanding and the timelike convergence condition (\ref{sec}) holds along the congruence,
\item the spatial scalar curvature is non-positive on average on 
$\Sigma$: $\left<\overline{R}\right>_{\Sigma}\leq 0$,
\item the cosmological constant is non-negative $\Lambda \geq 0$,
\item the energy-momentum tensor satisfies the dominant energy condition.
\end{enumerate}
If any single one of the following spatial averages
\begin{eqnarray*}
\Lambda , \,\, \left< \theta \right>_{\Sigma}, \,\, \left< \vartheta \right>_{\Sigma} ,
\,\, \left< \theta^2 \right>_{\Sigma}, \,\, \left< \vartheta^2 \right>_{\Sigma} ,\,\,
\left<K_{\mu\nu}K^{\mu\nu}\right>_{\Sigma},Ê\,\, \left<\overline{R}\right>_{\Sigma},\,\,
\left<v^{\mu}\nabla_{\mu}\vartheta\right>_{\Sigma},\\
\left<u^{\mu}\nabla_{\mu}\theta-\nabla_{\mu}a^{\mu}\right>_{\Sigma},\,\,\,
\left<T_{\mu\nu}e^{\mu}_{\alpha}e^{\nu}_{\beta}\right>_{\Sigma}, \hspace{3mm}
\left<R_{\mu\nu}e^{\mu}_{\alpha}e^{\nu}_{\beta}\right>_{\Sigma} \hspace{3mm}
\forall \alpha,\beta =0,1,2,3
\end{eqnarray*}
does \underline{not} vanish, then the spacetime is past timelike geodesically 
incomplete.
\label{th4}
\end{theo}
As before, notice that the first average in the second row can be replaced by 
$\left<u^{\mu}\nabla_{\mu}\theta-a_{\mu}a^{\mu}\right>_{\Sigma}$. Therefore, another sufficient condition is that $u^{\mu}\nabla_{\mu}\theta$ is non-positive everywhere but its average be non-zero.

\section{Conclusions}
Observe the following implication of the theorem. Under the
stated hypotheses, a non-vanishing average of {\em any} component
of the energy-momentum tensor (or of the Ricci tensor) ---such as the
energy density, the different pressures, the heat flux, etc.--- leads
to the existence of past singularities. It is quite
remarkable that one does not need to assume any specific type of
matter content (such as a perfect fluid, scalar field, ...); only the
physically compelling and well established dominant energy condition
is required. 
The theorem is valid for ``open" models, as the Cauchy hypersurface is
required to be non-compact. This is, however, no real restriction,
because for closed models there are stronger
resultsÊ\cite{HE,HP,P,Wald,S2}. As a matter of fact, closed expanding
non-singular models require the violation of the strong energy
condition \cite{S2}. There is no hope of physically acceptable closed
non-singular models satisfying the timelike convergence condition
(\ref{sec}) (without this condition, there are some examples
\cite{MS,S2}). 

Let me stress that the conclusion in theorem \ref{th4} is quite strong:
it tells us that the incompleteness is {\em to the past}. Besides, I believe that one can in fact prove a stronger theorem such that the geodesic timelike incompleteness is universal to the past. 
 
The main implication of the above results is this: a clear, decisive,
difference between singular and regular (globally hyperbolic)
expanding cosmological models is that the latter must have a vanishing
spatial average of the matter variables. Somehow, one could then say
that the regular models 
are {\em not} cosmological, if we believe that the Universe is
described by a more or less not too inhomogenous distribution of
matter.  
This is, on the whole, a very satisfactory result. 

\section*{Acknowledgments}
I am thankful to Naresh Dadhich and Leonardo Fern\'andez-Jambrina for many interesting conversations over the years. I am indebted to Matthias Blau and Harald Jerjen for pointing out a flaw in the proof of a previous version of the theorem. Financial support under
grants FIS2004-01626  and 
no. 9/UPV 00172.310-14456/2002 is acknowledged.

\section*{Appendix: the singularity-free model of \cite{S}}
In cylindrical coordinates $\{t,\rho,\varphi,z\}$ the line-element 
reads
\begin{eqnarray}
ds^2=\cosh^4(at)\cosh^2(3a\rho)(-dt^2+d\rho^2)+\hspace{2cm}\nonumber\\
\frac{1}{9a^2}\cosh^4(at)\cosh^{-2/3}(3a\rho)\sinh^2(3a\rho)d\varphi^2+
\cosh^{-2}(at)\cosh^{-2/3}(3a\rho)dz^2, \label{sol}
\end{eqnarray}
where $a>0$ is a constant. This is a {\em cylindrically symmetric} 
(the axis is defined by $\rho\rightarrow 0$) solution of the Einstein 
field equations (\ref{efe}) (with $\Lambda =0$ for simplicity)
for an energy-momentum tensor describing a perfect fluid:
$
T_{\mu\nu}=\varrho u_{\mu}u_{\nu}+p\, (g_{\mu\nu}+u_{\mu}u_{\nu}).
$
Here $\varrho$ is the energy density of the fluid given by
$$
\varrho = 15a^2\cosh^{-4}(at)\cosh^{-4}(3a\rho) \, ,
$$
and
$$
u_{\mu}=\left(-\cosh^2(at)\cosh(3a\rho),0,0,0\right)
$$
defines the unit velocity vector field of the fluid. Observe that 
$u^{\mu}$ is not geodesic (except at the axis), the acceleration 
field being
$$
a_{\mu}=\left(0,3a\tanh (3a\rho),0,0\right).
$$
The fluid has a {\em realistic} barotropic equation of state relating its isotropic pressure $p$ to $\varrho$ by:
$$
p=\frac{1}{3}\varrho.
$$
This is the canonical equation of state for radiation-dominated 
matter and is usually assumed to hold at early stages of the 
Universe. 
Note that the energy density and the pressure are regular everywhere, 
and 
one can in fact prove that the space-time (\ref{sol}) is completely 
free of 
singularities and geodesically complete \cite{CFS}.
For complete discussions on this spacetime see \cite{CFS} and 
sect.~7.6 in \cite{S2}. One can nevertheless see that the focusing effect on 
geodesics takes place fully in this space-time. This does not lead to any
problem with the existence of maximal geodesics between any 
pair of chronologically related points, see the discussion in \cite{S2}, pp.~829-830.

Space-time (\ref{sol}) satisfies the strongest causality condition: it 
is 
globally hyperbolic, any $t=$const.\ slice is a Cauchy hypersurface. 
All typical energy conditions, such as the dominant or the (strictly) 
strong ones (implying in particular (\ref{sec}) with the strict inequality) also hold everywhere. The 
fluid expansion is given by
\begin{equation}
\theta=\nabla_{\mu}u^{\mu} = 3a\frac{\sinh(at)}{\cosh^3(at)\cosh(3a\rho)}. 
\label{exp}
\end{equation}
Thus this universe is contracting for half of its history ($t<0$) 
and expanding for the second half ($t>0$), having a rebound at $t=0$ 
which is driven by the spatial gradient of pressure, or equivalently, by the acceleration $a_{\mu}$. Observe that the entire universe is expanding (that is, $\theta>0$) {\em 
everywhere} if $t>0$. Note that this is one of the assumptions in propositions \ref{th2}, \ref{th3} and theorem \ref{th4}. It is however obvious that, for any Cauchy hypersurface $\Sigma_T$ given by $t=T=$constant, the average $<\theta>_{\Sigma_T}=0$. As one can check for the explicit expression 
(\ref{exp}), $\theta$ is {\em strictly positive everywhere but not 
bounded from below by a positive constant} because 
$\lim_{\rho \rightarrow \infty}\theta=0$. Observe, however, that for finite $\rho$ one has $\lim_{z \rightarrow \infty}\theta >0$ and finite. 

Similarly, one can check that the scalar curvature of each $\Sigma_{T}$ is given by
$$
\overline R = 30a^2\cosh^{-4}(aT)\cosh^{-4}(3a\rho)>0
$$
which is positive everywhere. However $<\overline R >_{\Sigma_T}=0$,
and analogously $<\varrho>_{\Sigma_T}=<p>_{\Sigma_T}=0$. 
Observe also that
$$
a_{\mu}a^{\mu}=9a^2\frac{\sinh^2(3a\rho)}{\cosh^{4}(at)\cosh^{4}(3a\rho)}
$$
and thus $<a_{\mu}a^{\mu}>_{\Sigma_T}=0$. This implies that in this case
$<u^{\mu}\nabla_{\mu}\theta>_{\Sigma_T}=0$. The sign of $u^{\mu}\nabla_{\mu}\theta-\nabla_{\mu}a^{\mu}$ is negative everywhere, as can be easily checked:
$$
u^{\mu}\nabla_{\mu}\theta=3a^2\frac{1-3\tanh^2(at)}{\cosh^{4}(at)\cosh^{2}(3a\rho)},
\hspace{4mm}
\nabla_{\mu}a^{\mu}=3a^2\frac{\cosh^2(3a\rho)+5}{\cosh^{4}(at)\cosh^{4}(3a\rho)}.
$$
All this is in agreement with, and illustrates, the main theorem \ref{th4}, propositions \ref{th2} and \ref{th3}, and their corresponding remarks.

This simple model shows that there exist well-founded, well-behaved 
classical models which expand everywhere, satisfying all energy and 
causality conditions, and are singularity-free. However, as we have
just seen, the model is  
somehow not ``cosmological" to the extent that the above mentioned
spatial averages vanish.

\end{document}